\documentclass[twocolumn,amsmath,amssymb,eqsecnum,groupaddress]{revtex4-2}
\usepackage{graphicx,amsmath,relsize,epstopdf,upgreek,color,mathtools,bm,newtxtext,newtxmath,wasysym,rotating}
\usepackage[hyphenbreaks]{breakurl}
\usepackage[colorlinks=true,linkcolor=blue,citecolor=blue,urlcolor =blue]{hyperref}

\newcommand{\HG}{\mathop{\mathrm{HG}} \nolimits}
\newcommand{\Var}{\mathop{\mathrm{Var}} \nolimits}

\begin{document}

\title{Effects of coherence on temporal resolution}

\author{Syamsundar~De} 
\affiliation{Integrated Quantum Optics, Institute for Photonic Quantum Systems (PhoQS),  Warburger Stra\ss e 100, 33098 Paderborn, Germany}

\author{Jano~Gil-Lopez}
\affiliation{Integrated Quantum Optics, Institute for Photonic Quantum Systems (PhoQS),  Warburger Stra\ss e 100, 33098 Paderborn, Germany}

\author{Benjamin~Brecht}
\affiliation{Integrated Quantum Optics, Institute for Photonic Quantum Systems (PhoQS),  Warburger Stra\ss e 100, 33098 Paderborn, Germany}

\author{Christine~Silberhorn}
\affiliation{Integrated Quantum Optics, Institute for Photonic Quantum Systems (PhoQS),  Warburger Stra\ss e 100, 33098 Paderborn, Germany}

\author{Luis~L.~S\'{a}nchez-Soto}
\affiliation{Departamento de \'Optica, Facultad de F\'{\i}sica,
    Universidad Complutense, 28040 Madrid, Spain}
\affiliation{Max-Planck-Institut f\"ur die Physik des Lichts,
    Staudtstra\ss e 2, 91058 Erlangen, Germany}

\author{Zden\v{e}k~Hradil}
\affiliation{Department of Optics, Palack\'{y}  University, 17.~listopadu 12, 77146 Olomouc, Czech Republic} 

\author{Jaroslav~\v{R}eh\'a\v{c}ek}
\affiliation{Department of Optics, Palack\'{y}  University, 17.~listopadu 12, 77146 Olomouc, Czech Republic} 

\begin{abstract}
Measuring small separations between two optical sources, either in space or in time, constitute an important metrological challenge as standard intensity-only measurements fail for vanishing separations. Contrarily, it has been established that appropriate coherent mode projections can appraise arbitrarily small separations with quantum-limited precision. However, the question of whether the optical coherence brings any metrological advantage to mode projections is still a point of debate. Here, we elucidate this problem by experimentally investigating the effect of varying coherence on estimating the temporal separation between two single-photon pulses. We show that, for an accurate interpretation, special attention must be paid to properly normalize the quantum Fisher information to account for the strength of the signal.
\end{abstract}

\maketitle

\section{Introduction}

In numerous applications, including radar signal processing~\cite{Orlando:2017vj,Javadi:2020wx,Gini:2021to}, radio acoustic sounding~\cite{Peters:1983tn,Matuura:1986ut}, ultrasonic testing~\cite{Korte:1997aa}, and medical imaging~\cite{Drexler:2008aa}, one is faced with the challenge of determining the temporal delay between two closely spaced, overlapping, ultrashort pulses.  There are several efficient techniques to estimate these small offsets, such as the cross-correlation, phase-shift, and delay line methods~\cite{Dudacek:2015aa}. However, all of them conspicuously fail when the time delay is significantly shorter than the pulse duration.  
  
The same pitfall appears in the spatial domain: the resolution of an imaging system is limited by the size of its point spread function (PSF) that specifies the intensity response to a point source~\cite{Goodman:2004aa}. This gives an intuitive picture of the mechanisms that obstruct resolution, but it is very heuristic. For example, the Rayleigh limit~\cite{Rayleigh:1879aa} is defined as the distance from the center to the first minimum of the PSF. Yet that can be made arbitrarily small with ordinary linear optics, at the price of the side lobes becoming much higher than the central maximum.  This hints that estimating the separation between two points becomes also a matter of photon statistics~\cite{Schermelleh:2010aa}. 
    
Lately, the resolution limits have been revisited from the alternative perspective of quantum metrology.  The idea is to use the quantum Fisher information (FI) and the associated quantum Cram\'er-Rao bound (CRB) to assess how well the separation between two point sources can be estimated~\cite{Tsang:2016aa,Nair:2016aa,Lupo:2016aa,Tsang:2017aa,Zhou:2019aa,Peng:2021vu}. For direct imaging, the classical FI drops to zero as the separation between the sources decreases and the error with which we can determine the separation diverges accordingly, which has been dubbed Rayleigh’s curse~\cite{Tsang:2019aa}. Surprisingly, when the quantum FI (i.e., optimized over all measurements allowed by quantum mechanics) is calculated, it stays constant, evidencing that the Rayleigh limit is not essential. 
    
These remarkable predictions have fuelled a number of experimental implementations, both in the spatial~\cite{Paur:2016aa,Yang:2016aa,Tham:2017aa,Yang:2017aa} and the time-frequency~\cite{Donohue:2018aa} domains. The key behind these achievements is the use of phase-sensitive projections onto optimal modes~\cite{Rehacek:2017aa} instead of intensity measurements, as the latter discard the phase information carried by the signal.  
    
The approach has been generalized to more realistic scenarios, where the signals may have different intensities. This involves the simultaneous estimation of separation, centroid, and relative intensities~\cite{Rehacek:2017ab,Chrostowski:2017aa}.   Still in this multiparameter case~\cite{Ragy:2016aa,Szczykulska:2016aa,Albarelli:2020aa,Sidhu:2020aa}, optimal quantum-limited measurements have been worked out~\cite{Rehacek:2018aa} and experimentally demonstrated~\cite{Ansari:2021aa}.
    
The discussion thus far assumes incoherence between the signals.  This conforms with the conditions underlying Rayleigh's criterion. In the temporal domain, this happens with, for example, remote clocks (e.g., GPS)~\cite{Giorgetta:2013ti,Gao:2015uf,Kim:2020tv}, incoherent excitations in biological samples~\cite{Chen:2020vu}, condensed matter physics~\cite{Krausz:2009tf}, and astronomical observations~\cite{Krehlik:2017vi}.

A recent heated debate addressed the role of coherence in the resolution limits~\cite{Larson:2018aa,Tsang:2019ab,Larson:2019aa,Hradil:2019aa,Liang:2020aa,Wadood:2021aa,Tsang:2021aa,Liang:2021tw}. Any coherent superposition of two time-delayed pulses can be decomposed in terms of in-phase and anti-phase combinations of the two pulses. Yet, these two channels are not equivalent concerning the strength of the signal: the anti-phase mode does carry the information about the temporal separation, but the intensity in this mode vanishes as the time offset decreases. Hence, each photon therein carries a huge amount of information.

However, this effect is not necessarily a metrological advantage, because the input signal still contains many photons and thus, on average, the information per photon is limited. 

We stress that ignoring the resources required to generate the input signal might lead one to false conclusions about the information content of the measurement.  In this sense, the limit of incoherent mixtures represents a bound that cannot be overcome without prior information coded into the state~\cite{Hradil:2021vj}. Coherence may provide extra benefits by sorting information for different parameters into different channels.  We confirm here these predictions with an experiment that benefits from classical and quantum resources, both contained in the quantum FI of the signal. 

\section{Theoretical model} 

Let us first set the stage for our model.  To facilitate possible generalizations, we phrase what follows in a quantum language, so that a pulse waveform with complex temporal envelope $\psi( t )$ is  assigned to a ket $| \psi \rangle $, such that $\psi( t ) = \langle t | \psi \rangle$.  Here, we understand that an ideal (sharp) measurement of time would project on the state $ | t \rangle$ defined by a delta function in time or, in the frequency representation, by 
\begin{equation}
\langle \omega | t \rangle = \frac{1}{\sqrt{2 \pi}} e^{i \omega t} \, .
\end{equation}

We consider two pulses of identical shape but displaced by a time offset $\tau$, whose magnitude we want to estimate. We define the time-shifted versions, $\lvert \psi_{\pm} \rangle$, as $\psi_{\pm} (t) =  \psi(t \pm \tau/2)$. In addition, for convenience, we keep the total intensity normalized to unity: 
\begin{equation}
\int_{- \infty}^{\infty} dt \, 
[ \lvert \psi_{+} (t)\rvert^{2}  +  \lvert \psi_{-} (t)\rvert^{2}  ] = 1 \, . 
\end{equation}
In general,  $\langle \psi_{-} | \psi_{+} \rangle \neq 0$, so these modes are not orthogonal.  This overlap is at the heart of all the difficulties of the  problem, for it implies that the two modes cannot be separated by independent measurements.
    
To capture the essence of the problem we introduce symmetric and antisymmetric (nonnormalized) coherent modes  
\begin{align}
\label{superpos}
     \psi_{\mathrm{s}} (t) & =  \frac{1}{\sqrt{2}} 
     [ \psi_{+} (t)  + \psi_{-} (t) ] \,, \nonumber \\
     & \\
     \psi_{\mathrm{a}} (t) & = 
    \frac{1}{\sqrt{2}} [ \psi_{+} (t) - \psi_{-} (t) ] \, . \nonumber 
\end{align}
These coherent modes can be generated by a mode converter using linear optical transformations. Such an operation is, for instance, readily implemented by sending the two signals into different input ports of a balanced beam splitter. The total intensity is conserved in this process. Physically, the symmetric mode corresponds to an in-phase superposition of the time-shifted components, whereas the antisymmetric one corresponds to an anti-phase superposition.  
    
To proceed further, we need to specify the explicit waveform of the pulse. For simplicity, we will use the standard choice of a Gaussian profile 
\begin{equation}
\psi (t) = \frac{1}{(2 \pi \sigma_{t}^{2})^{1/4}} 
\exp \left ( - \frac{t^{2}}{4 \sigma_{t}^{2}} \right ) 
\end{equation} 
of width $\sigma_{t}$.

We first consider a fully incoherent mixture of the time-shifted components. Equivalently, this can be prepared as an incoherent sum of the in-phase and anti-phase channels; in the experimental realization this amounts to an equal mixing of the measurement data for each coherent mode in  post-processing. The situation can be thus represented by the following 
density matrix 
\begin{equation}
\varrho (\tau)  = \lvert \psi_{+} \rangle \langle \psi_{+} \rvert +  
\lvert \psi_{-}\rangle \langle \psi_{-} \rvert = 
\lvert \psi_{\mathrm{s}} \rangle  \langle \psi_{\mathrm{s}} \rvert  +  
\lvert \psi_{\mathrm{a}} \rangle \langle \psi_{\mathrm{a}}\rvert \, .
\end{equation} 
Now, we can directly apply quantum estimation theory.  The pivotal quantity is the quantum FI~\cite{Helstrom:1976ij}, which is a  mathematical measure of the sensitivity of an observable quantity (pulse waveform) to changes in its underlying parameters (time delay).  Replicating the calculations performed in the spatial domain~\cite{Paur:2016aa}, one immediately gets that the quantum FI, denoted by $\mathcal{Q}$, is constant, $\mathcal{Q} (\tau) = 1/(4\sigma^2_{t})$.  The associated quantum CRB ensures then that the variance of any unbiased estimator $\widehat{\tau}$ of the time delay $\tau$ is lower bounded by the reciprocal of the quantum FI (per single detection event); viz,
\begin{equation}
\Var_{\varrho} (\widehat{\tau}) \ge \frac{1}{\mathcal{Q} (\tau)} =  4  \sigma^2_{t} \, .
\end{equation} 

In what follows, it will prove convenient to look at the problem from a slightly different perspective. As the optimal measurement attaining the CRB is known, we can calculate the  FI for such a measurement.   In fact, the optimal scheme involves projections onto the successive derivatives (properly orthonormalized) of the pulse amplitude~\cite{Rehacek:2017aa}.  For our basic Gaussian waveform, this reduces to the Hermite-Gauss temporal modes  
\begin{align}
& \HG_{n} (t)  = \langle t | \HG_{n}\rangle  \nonumber \\
& = \frac{1}{(2\pi \sigma_{t}^{2})^{1/4}} \frac{1}{(2^{n} \, n!)^{1/2}} 
H_{n} \left ( \frac{t}{\sqrt{2} \sigma_{t}} \right )\,  \exp \left ( - \frac{t^{2}}{4 \sigma_{t}^{2}}\right ) .
\end{align}
    
Then, we have the following detection probabilities
    \begin{eqnarray}
    p_{s} (n | \tau ) \equiv |\langle \HG_{n} | \psi_{\mathrm{s}} \rangle|^2
    =\left\{
    \begin{array}{ll}
    p_{n} ( \tau)  \quad \qquad   & n=0,2,4,\ldots\\
    & \\
    0 & n=1,3,5,\ldots
    \end{array}
    \right. \nonumber \\
    \\
    p_{a} (n | \tau ) \equiv|\langle \HG_{n}| \psi_{\mathrm{a}} \rangle|^2
    =\left\{
    \begin{array}{ll}
    0 \quad \quad & n=0,2,4,\ldots\\
    & \\
    p_{n} ( \tau) \quad \qquad & n=1,3,5,\ldots
    \end{array}
    \right. \nonumber
    \end{eqnarray}
Here, $p_{\alpha} (n | \tau )$ ($\alpha \in \{a, s\}$)  denotes the probability density for a detection when projecting the symmetric (anti-symmetric) coherent mode $| \psi_\mathrm{s} \rangle$ ($|\psi_\mathrm{a}\rangle$) onto the mode $\HG_{n}$, conditional on the value of the time delay $\tau$ and we have defined 
\begin{equation}
p_{n} (\tau) = \frac{1}{n! \, 16^{n}} \left ( \frac{\tau}{\sigma_{t}} 
\right )^{2n} \exp \left ( - \frac{\tau^{2}}{16 \sigma_{t}^{2}} \right ) \, .
\end{equation} 

Due to the limitations of the experimental setup described later, we cannot generate incoherent signals directly. However, we can generate coherent in-phase and anti-phase superpositions, which directly correspond to the output ports of the beam splitter (\ref{superpos}). Mixing the measurement data for both superpositions in post-processing allows for realizing an arbitrary amount of coherence between the two signal pulses. 

Note that measuring the output of the interference between two signals is relevant for many applications, such as stellar interferometry, and thus not just a convenient theoretical approach. In consequence, we have now 
\begin{equation}
p_{\mathrm{incoh}} (n | \tau) = 
p_{s} (n | \tau ) + p_{a} (n | \tau ) =   p_{n} ( \tau ) \, .
\end{equation}

The classical FI about $\tau$ from these mode projections on in-phase and anti-phase states is
\begin{equation}
    \label{eq:FIinc}
F_{\mathrm{incoh}} (\tau) =  \sum_{n} \frac{1}{p_{\mathrm{incoh}} (n | \tau)} 
\left [ \frac{\partial p_{\mathrm{incoh}} (n | \tau)}{\partial n} \right ]^2  \, .
\end{equation}
Since in-phase and anti-phase detection  happen in even and odd mode projections, respectively, no information is lost in this process and the incoherent FI does saturate the quantum bound: 
\begin{equation}
F_{\mathrm{incoh}} (\tau) = \frac{1}{4\sigma^2_{t}} \, .
\end{equation} 

The incoherent mixture is given  either by a sum of time-delayed Gauss components or by a sum of unnormalized in-phase and anti-phase superpositions as in \eqref{superpos}. The FI for the  sum of probabilities in $p_{\mathrm{incoh}}$ saturates the ultimate limit for time-localization of input components $\psi_{\pm}$ and the beam splitter action \eqref{superpos} is a unitary process, preserving information. The Hermite-Gauss projections do saturate the quantum bound simultaneously for both in-phase and anti-phase superpositions. 

Let us now consider the opposite case of fully coherent signals. The estimation of the time shift $\tau$ requires projections applied to in-phase and anti-phase states; the results are
\begin{eqnarray}
\label{fishinfo}
    F_{s} (\tau) & = &  \frac{1}{8 \sigma_{t}^2} - \left ( \frac{1}{8 \sigma_{t}^2} - 
    \frac{\tau^{2}}{32 \sigma_{t}^{4}} \right ) 
    \exp \left ( - \frac{\tau^{2}}{8 \sigma_{t}^{2}} \right ) \, , \nonumber \\
    & & \label{eq:Fas} \\
    F_{a} (\tau) & = &  \frac{1}{8 \sigma_{t}^2} + \left ( \frac{1}{8 \sigma_{t}^2} - 
    \frac{\tau^{2}}{32 \sigma_{t}^{4}} \right ) 
    \exp \left ( - \frac{\tau^{2}}{8 \sigma_{t}^{2}} \right ) \, . \nonumber
\end{eqnarray}
If we detect both outputs, we have 
\begin{equation}
F_{\mathrm{coh}} (\tau ) = \frac{1}{4 \sigma^{2}_{t}} \, ,
\end{equation} 
which also saturates the quantum bound.  Note, though, that for small separations, $\tau\rightarrow 0$, we have 
\begin{equation} 
F_{s} (\tau)  \simeq 0\, , \qquad 
F_{a} (\tau)  \simeq  \frac{1}{4 \sigma^{2}_{t}} \, ,
\end{equation} 
that is, almost all the information resides in the anti-phase channel, $F_{a}$. However, in this limit the intensity available in this channel becomes $\sum_n p_{a} (n | \tau) \rightarrow 0$. Nonetheless, we must remember that a constant amount of input intensity is spent on generating the anti-phase superposition for any separation. In a sense, the beam splitter acts as an information sorter that directs the information about the timing separation to the weak anti-phase channel. The majority of the signal intensity is sent to the in-phase output, where other parameters (e.g. the timing centroid) can be simultaneously accessed.  In the case of complete incoherence, no interference on the beam splitter occurs: the measured intensity in the anti-phase output is half of the input intensity, regardless of the timing separation. In this case, a simple intensity measurement is insufficient to resolve the separation of the two signals. It is known, however, that mode projections are ideal and, in fact, remain optimal for any degree of coherence.

A PSF-independent formulation can be provided by defining a modified quantum FI for states with parameter-dependent norm $ N ( \tau)= \langle \psi (\tau) | \psi (\tau)  \rangle$;  it reads~\cite{Hradil:2021vj}
\begin{align}
\widetilde{\mathcal{Q}} (\tau) & =   4\langle \partial_{\tau} \psi (\tau) | \partial_{\tau} \psi (\tau) \rangle   \nonumber \\
& +   \frac{1}{N(\tau)} \left [\langle \psi (\tau) | \partial_{\tau} \psi (\tau) \rangle  -  
\langle \partial_\tau  \psi (\tau) |  \psi(\tau)\rangle  \right ]^{2} \, .
\end{align}
Applying this to the superpositions (\ref{superpos}) confirms the optimality of Hermite-Gauss projections.

As we said before, partial coherence has been the subject of a recent controversy~~\cite{Larson:2018aa,Tsang:2019ab,Larson:2019aa,Hradil:2019aa,Liang:2020aa,Wadood:2021aa}. Actually, we maintain that partial coherence just redistributes the information into two (or more) interfering channels. Information is carried both by the norm (i.e., intensity-modulated by the estimated parameter) and the underlying normalized quantum state. However total information is preserved and can be extracted with a suitable measurement.

For all degrees of coherence and to remove the impact of an intensity-dependent norm in the experiment, we realize partially coherent states as mixtures of two coherent detection schemes with $s$ and $a$ channels interchanged.  We quantify coherence with the parameter $\gamma$ of such convex combinations, where $\gamma=0$ means fully coherent and $\gamma=1/2$ means fully incoherent. The projection on the same temporal Hermite-Gauss modes of in-phase and anti-phase channels are
\begin{eqnarray}
\label{eq:mixing}
    {P}_{s}^{\gamma} (n | \tau)
    =\left\{
    \begin{array}{ll}
    (1-\gamma) p (n | \tau ) \quad \qquad& n=0,2,4,\ldots\\
    & \\
    \gamma \, p (n | \tau ) \quad & n=1,3,5,\ldots
    \end{array}
    \right . \nonumber \\
    \\
    {P}_{a}^{\gamma} (n | \tau)
    =\left\{
    \begin{array}{ll}
    \gamma \, p (n | \tau) \quad \qquad& n=0,2,4,\ldots\\
    & \\
    (1-\gamma) p (n | \tau) \quad \qquad & n=1,3,5,\ldots
    \end{array}
    \right. \nonumber
    \end{eqnarray}
Notice that for $\gamma=1/2$, we obtain two identical sets of probabilities ${P}^{1/2}_{a}(n | \tau) = {P}^{1/2}_{s}(n | \tau)$, which upon adding  we recover the incoherent probabilities $p_{\mathrm{incoh}} (n| \tau)$. Naturally, the total FI again saturates the quantum bound for any $\gamma$ and all separations
    \begin{eqnarray}
    {F}_{s}^{\gamma} (\tau ) & = & (1-\gamma) F_{s} (\tau) + \gamma F_{a} (\tau )\,, 
    \nonumber \\
    & & \\ 
    {F}_{a}^{\gamma} (\tau) & = & \gamma F_{s} (\tau) + (1-\gamma) F_{a} (\tau) \, ,
    \nonumber
    \end{eqnarray}
and consequently 
\begin{equation}
{F}^{\gamma} (\tau ) = {F}_{s}^{\gamma} (\tau ) + {F}_{a}^{\gamma} (\tau ) = F_{\mathrm{incoh}} (\tau) = \frac{1}{4\sigma_{t}^2} \, .
\end{equation} 

From the discussion thus far, it should be clear that Hermite-Gauss temporal modes  are optimal for \textit{any} degree of coherence. Of course, any other complete mode decomposition with even/odd temporal symmetry will do the same job~\cite{Rehacek:2017aa}. However, these projections require sophisticated equipment. Intensity detection  is still the cut-and-dried method used in the laboratory. As the dominant part of the information about separation is contained in the norm of the anti-phase superposition, total intensity represents a valuable source of information. 
  
For incoherent states, intensity detection leads to Rayleigh's curse and it is not optimal. For full coherence, however, intensity detection is one optimal solution. This can be readily shown by calculating the FI for the intensity profiles of in-phase and anti-phase channels, whose probabilities of detection are $P_{\alpha}(t | \tau )= \lvert \psi_{\alpha} (t) \rvert^2$, with $\alpha \in \{a, s \}$, whence we get
\begin{equation}
    F_{\alpha}^{\mathrm{int}} (\tau) = \int dt \frac{1}{P_{\alpha}(t|\tau)} 
    \left [ \frac{\partial P_{\alpha}(t|\tau)}{\partial \tau} \right ]^2 = 
    F_{\alpha} (\tau) \, ,
\end{equation}
with $F_{\alpha} (\tau)$ given by \eqref{eq:Fas}. Therefore, $F_{s}^{\mathrm{int}} (\tau) + F_{a}^{\mathrm{int}} (\tau)= 1/\sigma_{t}$ and there is no need for sophisticated mode projections when working with a fully coherent signal. However, the temporal resolution can be strongly improved by these detections for partially coherent and incoherent signals, especially in the limit $\tau \rightarrow 0$, which is precisely the regime of interest.

\section{Experimental results}

\begin{figure}[t]
    \centering
    \includegraphics[width=\columnwidth]{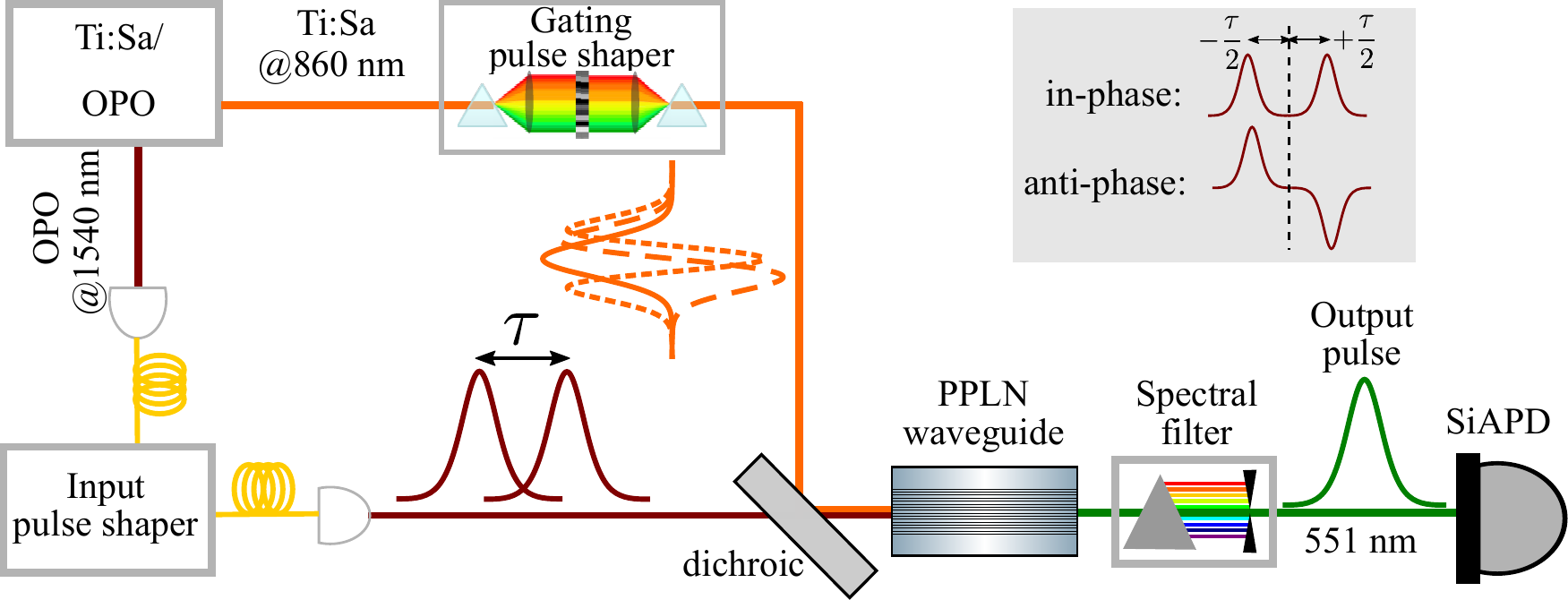}
    \caption{Schematic of the experimental setup.  Inset: in-phase and anti-phase input signals.    The in-phase and anti-phase input pulses with different time shifts ($\tau$) are derived from a broadband OPO at 1540 nm and attenuated to the few-photon levels using a commercial pulse shaper.   Gating pulses at 860 nm with Hermite-Gauss profiles are shaped with an in-house pulse shaper.   The input and gating pulses co-propagate through a PPLN waveguide. A sum-frequency process generates green output photons  at 551~nm which are bandpass filtered and subsequently counted using a silicon avalanche photo-diode (SiAPD).} 
    \label{fig:setup}
\end{figure}
The key building block in our experiment for the implementation of the optimal Hermite-Gauss temporal-mode projections is a quantum pulse gate (QPG)~\cite{Eckstein:2011vg,Brecht:2014vl}. It is based on group-velocity matched sum-frequency generation between a strong gating pulse and a weak signal pulse in a nonlinear waveguide. Detecting the upconverted photons then realizes  projective measurements, in which the temporal-mode projections are defined by the shapes of the gating pulse. 
        
The detailed scheme of our experimental setup is sketched in Fig.~\ref{fig:setup}.  A titanium-sapphire (Ti:Sapph) laser and an optical parametric oscillator (OPO) provide the gating and the input pulses with a repetition rate of $80~\mathrm{ MHz}$, respectively. The gating pulses are carved from a laser spectrum centered at $860~\mathrm{nm}$ with a full-width at half-maximum (FWHM) bandwidth of $7.25~\mathrm{nm}$. The gating pulses are shaped into user-defined Hermite-Gauss temporal modes with a home-built pulse shaper, comprised of a spatial light modulator at the Fourier plane of a $4f$ line arrangement.

The input pulses are derived from the OPO, delivering light at $1540~\mathrm{nm}$ with an FWHM bandwidth of $23~\mathrm{nm}$. A commercial fiber-coupled pulse shaper prepares the input signal that consists of coherent superpositions of two time-shifted Gaussian pulses of $1.26~\mathrm{ ps}$ width with equal (in-phase) or opposite (anti-phase) phase. As shown in the inset of Fig.~\ref{fig:setup}, one receives a positive time shift of $\tau/2$ and the other one receives a negative time shift of $-\tau/2$ with respect to a reference that is set at zero without loss of generality.    In our experiment, seven different time shifts ranging from $0$ to $\sigma_{\tau}$ are realized for both in-phase and anti-phase inputs by programming the pulse shaper.   Moreover, the input pulses are attenuated to a few photons per pulse. 
        
The shaped gating and the input pulses are combined on a dichroic mirror and then coupled into a home-built QPG---a $3.5~\mathrm{cm}$ long periodically poled lithium niobate (PPLN) waveguide with a poling period of $4.4~\mu\mathrm{m}$.   The waveguide is designed for spatially single-mode propagation of the input signal, whereas the propagation of the gating beam in the fundamental spatial mode is ensured via optical mode matching.   The sum-frequency process yields a green output at $551~\mathrm{nm}$ that undergoes tight spectral filtering in a 4-f line to discard the phase-matching side-lobes, resulting in an FWHM bandwidth of $40~\mathrm{pm}$.  The filtered output is detected with a fiber-coupled silicon avalanche photodiode (SiAPD). Finally, we record the single counts using a commercial time-tagger. We record data for $16~\textrm{ms}$ for each setting of the input and gating pulses, and repeat the measurement 100 times for the statistical analysis of the data. 
    \begin{figure}[t]
    \includegraphics[width=0.85\columnwidth]{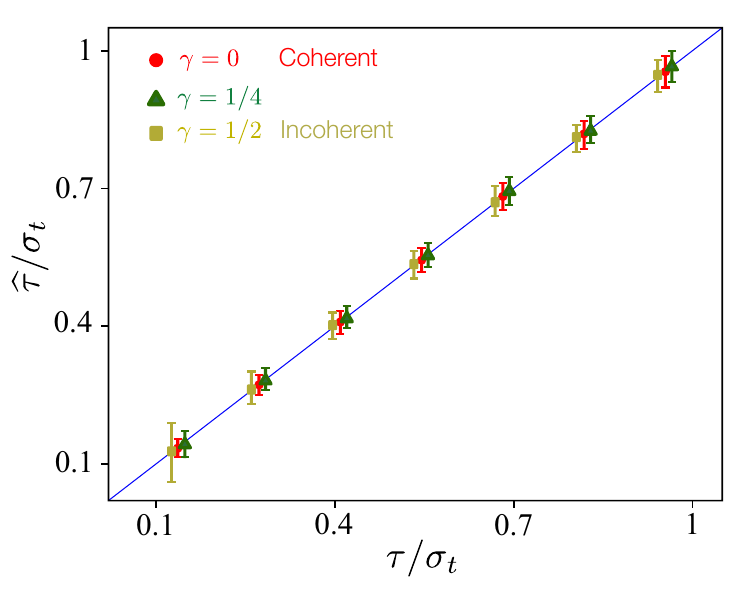}
    \caption{Estimates $\widehat{\tau}$ versus true separations $\tau$, both in units of the pulse width $\sigma_{t}$ for the three values of the coherence indicated in the inset.}
    \label{fig:2}
    \end{figure}

In our experiment, different coherence strengths are achieved in post-processing by controlled mixing of the measured data for the in-phase and the anti-phase input signals---each of them separately corresponds to the fully coherent case, their equal mixing leads to the incoherent case, and  unequal mixing corresponds to the partial coherence.  Rather than using the theoretical projections, we determine the actual input-output relations of the implemented imperfect QPG device, enabling the construction of an unbiased estimator of the time separation despite the limited selectivity of the device~\cite{Donohue:2018aa}. 

For each coherence setting, this is achieved by fitting the average responses of the first four Hermite-Gauss projections with fourth-order polynomials in $\tau$. The resulting measurement matrix is then used to process the individual detections taking the generalized least squares estimator constrained by the condition $\widehat{\tau} \ge 0$. Estimator statistics are calculated from $100$ repeated measurements for each combination of $\tau$ and $\gamma$ parameters. 

    \begin{figure}[t]
    \includegraphics[width=0.95\columnwidth]{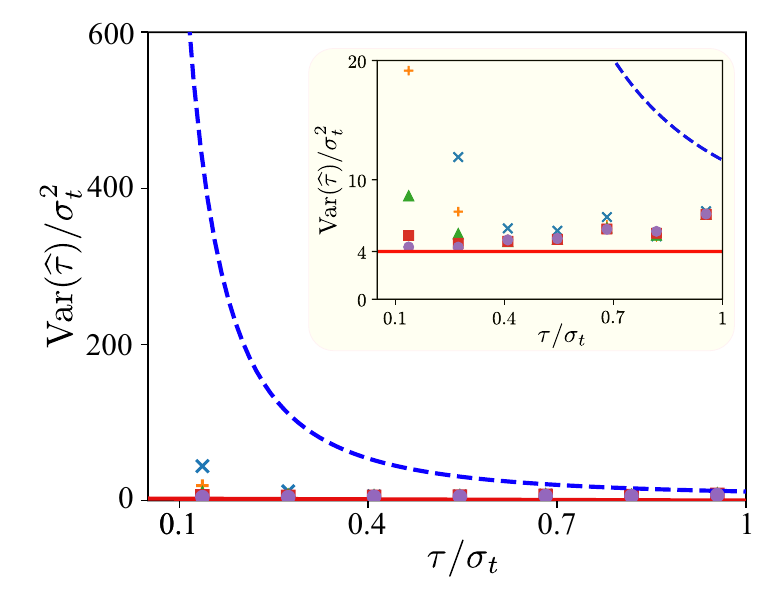}
    \caption{Variance of the estimator $\widehat{\tau}$ as a function of the true time offset $\tau$ for several values of the coherence parameter $\gamma$:  1/2 (blue crosses), 3/8 (orange crosses), 1/4 (green triangles), 1/8 (red squares), and 0 (purple circles). The ultimate limit given by the quantum CRB is given by the red solid line, whereas the classical incoherent detection limit is the broken blue line. The inset shows a zoomed version where we can better appreciate the behavior.}
    \label{fig:3}
    \end{figure}

We had to take special care to avoid unwanted temporal drift between the signal and gating fields that mainly originates from thermal fluctuations. To remove the effect of residual drift, the two fields are recentered by software control after every 10 measurement runs of each setting. 
    
Figure~\ref{fig:2} shows the statistics of the experimental estimates of $\tau$ for incoherent, coherent, and partially coherent superpositions. Mean values are plotted with standard deviation bars around. The true values are inside the standard deviation intervals for all separations and the estimator bias is negligible.

Figure~\ref{fig:3} shows the estimation errors (quantified by the variance) for five different mixtures ranging from fully coherent to incoherent. Coherent estimates saturate the quantum bound for small separations: we experimentally resolve temporal offsets ten times smaller than their pulse duration, with a tenfold improvement in precision over the intensity-only CRB.  When coherence is reduced, we see an increase of the experimentally determined variances, especially for the smallest measured time separations. This effect is a consequence of a tiny, yet nonnegligible, crosstalk between the odd and even Hermite-Gauss projections. Leakage of the strongly populated $\HG_{0}$ mode of the symmetric channel towards odd projections, upon mixing the channels as in \eqref{eq:mixing}, degrades the information carried by the weaker odd modes (and, in particular, the mode $\HG_{1}$) of the antisymmetric channel. But even with these imperfections, we are still much below the intensity-only CRB.

    \begin{figure}[t]
    \includegraphics[width=0.95\columnwidth]{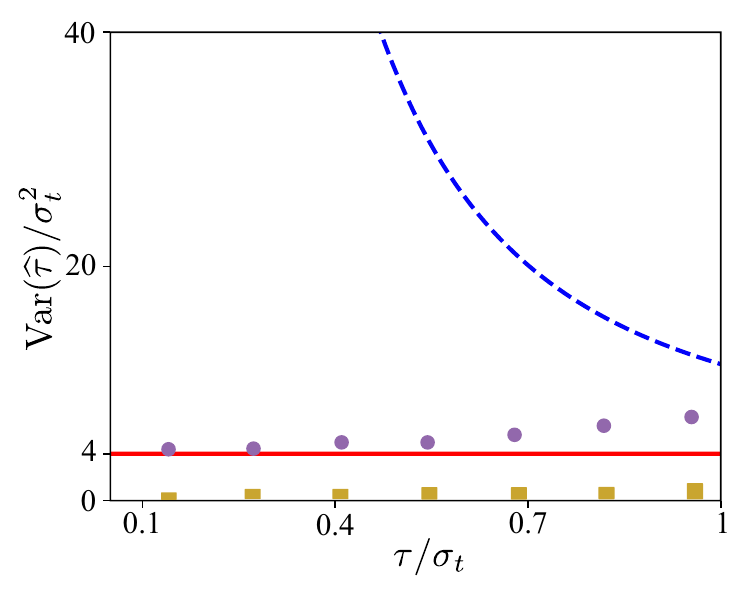}
    \caption{Estimation errors for fully coherent signal ($\gamma=0$) per single total detection (purple circles, as in Fig.~\ref{fig:3}) and per single antisymmetric ($\psi_a$) detection (brown squares). The  quantum (red solid line) and classical incoherent (blue broken line) limits are also shown.} 
    \label{fig:4}
    \end{figure}

Finally, in Fig.~\ref{fig:4} we show the same errors for fully coherent superposition both per single total detection and per single detection in the antisymmetric ($\psi_a$) channel.  The information carried by one such copy seemingly diverges at $\tau \rightarrow  0$  and violates the incoherent QCRLB (brown squares). However, a proper resource counting, in this case the total single counts, leads to the correct quantum-limited estimation error (purple circles). This is another way of expressing our previous statement that coherence acts like an information sorter condensing information about separation in the \textit{dark} antisymmetric channel.

\section{Concluding remarks}

By resorting to mode-selective measurements, sub-pulse-width separations can be estimated with quantum-limited precision for a full range of temporal coherence.   Seemingly, coherence by itself does not provide a direct metrological advantage in time resolution; incoherent superpositions setting the ultimate limits in all cases. 

However, we stress that coherence can be exploited as an information sorter, distributing information about different parameters into different channels. In our case, for small-time delays, all information is accessible from vanishing intensity in the anti-symmetric channel while the bulk of the intensity goes into the symmetric channel, left available for measurements of other relevant physical parameters.
   
\section*{Acknowledgments}

We thank V. Ansari and J. M. Donohue for discussions. We acknowledge the European Union's Horizon 2020 research and innovation program (Grants ApresSF and STORMYTUNE), Deutsche Forschungsgemeinschaft (231447078--TRR 142), the Grant Agency of the Czech Republic (18-04291S), and the Spanish Ministerio de Ciencia e Innovaci\'on (PGC2018-099183-B-I00).  


%

\end{document}